\documentclass[12pt,leqno]{article} 
\usepackage{amssymb, amsthm}
\usepackage{amsmath}
\usepackage{mathtools}
\usepackage{epsfig}
\usepackage{mathrsfs}
\usepackage{color,subfigure}
\newcommand{\ds}{\displaystyle}
\newcommand{\lf}{\left}
\newcommand{\rt}{\right}

\newcommand{\tit}{\textit}
\newcommand{\tbf}{\textbf}

\newcommand{\ovl}{\overline}

\newcommand{\ndconst}{\text{const }}

\newcommand{\noi}{\noindent}

\newcommand{\beq}{\begin{equation}}
\newcommand{\eeq}{\end{equation}}

\newcommand{\CC}{\mathbb{C}}
\newcommand{\RR}{\mathbb{R}}
\newcommand{\ZZ}{\mathbb{Z}}
\newcommand{\mA}{\mathcal{A}}
\newcommand{\mL}{\mathcal{L}}
\newcommand{\msE}{\mathcal{E}}
\newcommand{\msC}{\mathscr{C}}
\newcommand{\dO}{\dot{O}}
\newcommand{\dM}{\dot{M}}
\newcommand{\dx}{\dot{x}}
\newcommand{\dy}{\dot{y}}
\newcommand{\vom}{\vec{\om}}
\newcommand{\pu}{u^\#}
\newcommand{\px}{x^\#}
\newcommand{\py}{y^\#}

\newcommand{\tH}{\tilde{H}}
\newcommand{\tT}{\tilde{T}}
\newcommand{\tq}{\tilde{q}}
\newcommand{\hH}{\widehat{H}}
\newcommand{\hU}{\widehat{U}}
\newcommand{\hk}{\hat{k}}
\newcommand{\hr}{\hat{r}}
\newcommand{\hx}{\hat{x}}
\newcommand{\hy}{\hat{y}}
\newcommand{\hz}{\hat{z}}
\newcommand{\hlm}{\hat{\la}}

\newcommand{\al}{\alpha}

\newcommand{\de}{\delta}
\newcommand{\De}{\Delta}
\newcommand{\ep}{\epsilon}
\newcommand{\om}{\omega}
\newcommand{\Om}{\Omega}
\newcommand{\nab}{\nabla}
\newcommand{\sg}{\sigma}
\newcommand{\Sg}{\Sigma}
\newcommand{\la}{\lambda}

\newcommand{\p}{\partial}

\newcommand{\half}{\frac{1}{2}}

\theoremstyle{plain}
\newtheorem*{theorem}{Theorem}

\theoremstyle{definition}
\newtheorem*{notes}{Notes}

\newtheorem*{question}{Question}

\begin{document}

\title{Coxeter Lecture Series\\
Fields Institute\\
August 8-10, 2017}
\date{\ }
\maketitle

\newpage

Three lectures on 
\\
\centerline{``Fifty Years of KdV:  An Integrable System''}
\\
\\[-2mm]

\centerline{Percy Deift\footnote{The work of the author was supported in part
by NSF Grant DMS--1300965}}
\centerline{Courant Institute, NYU}
\smallskip

The goal in the first two Coxeter lectures was to give an answer to the question
\\
\centerline{``What is an integrable system?''}

\noindent
and to describe some of the tools that are available to identify and integrate
such systems.  The goal of the third lecture was to describe the role of 
integrable systems in certain numerical computations, particularly the 
computation of the eigenvalues of a matrix.  This paper closely follows
these three Coxeter lectures, and is written in an informal style
with an abbreviated list of references.
Detailed and more extensive references are readily available on the web.  The
list of authors mentioned is not meant in any way to be a detailed historical
account of the development of the field and I ask the reader for his'r 
indulgence on this score.

The notion of an integrable system originates in the attempts in the 17$^{th}$
and 18$^{th}$ centuries to integrate certain specific dynamical systems in 
some explicit way.  Implicit in the notion is that the integration reveals
the long-time behavior of the system at hand.  The seminal event in these 
developments was Newton's solution of the two-body problem, which verified
Kepler's laws, and by the end of the 19$^{th}$ century many dynamical systems
of great interest had been integrated, including classical spinning tops, 
geodesic flow on an ellipsoid, the Neumann problem for constrained harmonic
oscillators, and perhaps most spectacularly, Kowalewski's spinning
top.  In the 19$^{th}$ century, the general and very useful notion of 
\tit{Liouville integrability} for Hamiltonian systems, was introduced: If
a Hamiltonian system with Hamiltonian $H$ and $n$ degrees of freedom has
$n$ independent, Poisson commuting integrals, $I_1, \dots, I_n$, then
the flow $t\mapsto z(t)$ generated by $H$ can be integrated explicitly by 
quadrature, or symbolically,
\beq\label{eq1}
\begin{cases}
I_k\lf(z(t)\rt) &= \ndconst,\; 1 \le k \le n, \; \text{rank }\lf(dI_1, \dots, 
dI_n\rt)=n, \; \lf\{I_k, I_j\rt\}=0, \quad 1\le j, \;k\le n\\
&\quad \Rightarrow \text{ explicit integration}.
\end{cases}
\eeq
Around the same time the Hamilton-Jacobi equation was introduced,
which proved to be equally useful in integrating systems.

The modern theory of integrable systems began in 1967 with 
the discovery by Gardner, Greene, Kruskal and Miura [GGKM] of a method
to solve the Korteweg de Vries (KdV) equation 
\begin{align}\label{eq2}
&q_t + 6qq_x - q_{xxx} =0\\
&q(x,t)_{t=0} = q_0 (x) \to 0 \quad\text{as}\quad |x|\to \infty\ . \nonumber 
\end{align}
The method was very remarkable and highly original and expressed
the solution of KdV in terms of the spectral and scattering theory
of the Schr\"odinger operator $L(t) = -\p^2_x + q(x,t)$, acting in 
$L^2 (-\infty < x < \infty)$ for each $t$.  In 1968 Peter Lax \cite{Lax}
reformulated \cite{GGKM} in the following way.  For $L(t)= - \p^2_x + 
q(x,t)$ and $B(t)\equiv 4 \p^3_x -6q \, \p_x - 3q_x$.
\begin{align}\label{eq3}
&\text{KdV } \equiv \p_t \, L = [B, L]= BL-LB\\
&\qquad\; \equiv \text{ isospectral deformation of $L(t)$}\nonumber\\[2mm]
&\Rightarrow \text{ spec $(L(t))$ = spec $(L(0))$} \Rightarrow 
\text{  integrals of the motion for KdV}. \nonumber
\end{align}
$L, B$ are called \tit{Lax pairs}: By the 1970's, Lax pairs for the Nonlinear
Schr\"odinger Equation (NLS), the Sine-Gordon equation, the Toda lattice, 
\dots, had been found, and these systems had been integrated as in the 
case of KdV in terms of the spectral and scattering theory of their 
associated ``L'' operators.

Over the years there have been many ideas and much discussion
of what it means for a system to be integrable, i.e.\ explicitly solvable.
Is a Hamiltonian system with $n$ degrees of freedom integrable 
if and only if the system is Liouville integrable, i.e.\ the system
has $n$ independent, commuting integrals?  Certainly as explained above,
Liouville integrability implies explicit solvability.  But is the
converse true?  If we can solve the system in some explicit fashion, 
is it necessarily Liouville integrable?  We will say more about this matter
further on.  Is a system integrable if and only if it has a 
Lax pair representation as in (3)?  There is, however, a problem
with the Lax-pair approach from the get-go.  For example, if we are
investigating a flow on $n\times n$ matrices, then  a Lax-pair would 
guarantee at most $n$ integrals, viz., the eigenvalues, whereas an $n\times n$
system has $O(n^2)$ degrees of freedom --- too little, a priori, for 
Liouville integrability.  The situation is in fact even more complicated.
Indeed, suppose we are investigating a flow on real skew-symmetric
$n\times n$ matrices $A$ --- i.e.\ a flow for a generalized top.  Such 
matrices constitute the dual Lie algebra of the orthogonal group $O_n$,
and so carry a natural Lie-Poisson structure.  The symplectic leaves
of this structure are the co-adjoint orbits of $O_n$
\beq\label{eq4}
\mA = \mA_A = \lf\{ O\;A\; O^{T}: O\;\in\; O_n\rt\}
\eeq
Thus \tbf{any} Hamiltonian flow $t\to A(t)$ on $\mA, \; A(t=0)=A$,
must have the form 
\beq\label{eq5}
A(t) = O(t)\; A\;O(t)^{\,T}
\eeq
for some $O(t)\in \mA$ and hence has Lax-pair form
\begin{align}
\frac{dA}{dt} &= \dO\;A\;O^{T} + O\:A\:\dO^{T} = [B,A] \label{eq6} \\
\intertext{where}
B &= \dO \: O^{T} =-B^T \label{eq7}
\end{align}
The Lax-pair form guarantees that the eigenvalues $\{\la_i\}$ of $A$
are constants of the motion.  But we see from (4) that the 
co-adjoint orbit through $A$ is simply specified by the eigenvalues of $A$. 
In other words the eigenvalues of $A$ are just parameters for the 
symplectic leaves under considerations: They are of no help in 
integrating the system:  Indeed $d\la_i |_{\mA_A} =0$ for all $i$.
So for a Lax-pair formulation to be useful, we need
\beq\label{eq8}
\text{Lax pair} + \text{ ``something''}
\eeq
So, what is the ``something''?  A Lax-pair is a proclamation, a 
marker, as it were, on a treasure map that says ``Look here!''
The real challenge in each case is to turn the Lax-pair, if possible,
into an effective tool to solve the equation.  In other words, the
real task is to find the ``something'' to dig up the treasure!
Perhaps the best description of Lax-pairs is a restatement of Yogi Berra's
famous dictum ``If you come to a fork in the road, take it''.  So if you come
upon a Lax-pair, take it!

Over the years, with ideas going back and forth, Liouville integrability,
Lax-pairs, ``algebraic integrability'', ``monodromy'', the discussion of
what is an integrable system has been at times, shall we say, 
quite lively.  There is, for example, the story of Henry McKean
and Herman Flashka discussing integrability, when one of them, and
I'm not sure which one, said to the other:  ``So you want to know what is
an integrable system?  I'll tell you!  You didn't think I could solve it.
But I can!''

In this ``wild west'' spirit, many developments were taking place in 
integrable systems.  What was not at all clear at the time, 
however, was that these developments would provide tools to analyze
mathematical and physical problems in areas \tbf{far removed from 
their original dynamical origin}.  These tools constitute what may now be 
viewed as an \tbf{integrable method (IM)}.

There is a picture that I like that illustrates, very schematically,
the intersection of IM with different areas of mathematics.  
Imagine some high dimensional space, the ``space of problems''.  
The space contains a 
large number of ``parallel'' planes, stacked one on top of the other and 
separated.  The planes are labeled as follows:  dynamical systems, 
probability theory and statistical mechanics, geometry, combinatorics,
statistical mechanics, classical analysis, numerical analysis, representation
theory, algebraic geometry, transportation theory, \dots.  In addition, there
is another plane in the space labeled ``the integrable method (IM)'':
Any problem lying on IM can be solved/integrated by tools taken from the 
integrable method.  Now the fact of the matter is that the IM-plane intersects
all of the parallel planes described above:  Problems lying on the intersection
of any one of these planes with the IM-plane are thus solvable by the 
integrable method.  

\begin{figure}[tbp]\label{fig1}
  \centering
  \includegraphics[width=.5\linewidth]{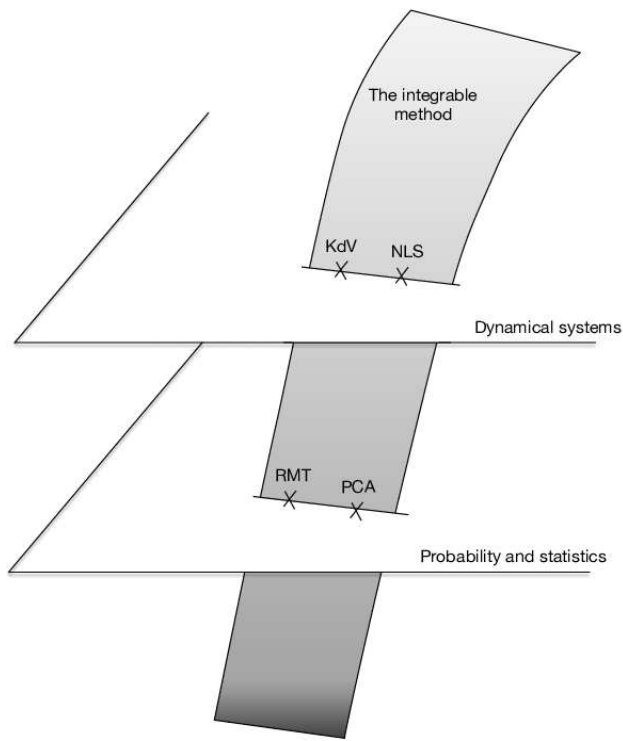}
\caption{Intersections of the Integrable Method }  
\end{figure}

\newpage
\noi
For each parallel plane we have, for example, the 
following intersection points:
\begin{itemize}
\item dynamical systems:  Korteweg-de Vries (KdV), Nonlinear Schr\"odinger
(NLS), Toda, Sine-Gordon, \dots 
\item probability theory and statistics:  Random matrix theory (RMT), 
Integrable probability theory, Principal component analysis (PCA), \dots
\item geometry:  spaces of constant negative curvature $R$, general relativity 
in $1+1$ dimensions, \dots 
\item combinatorics: Ulam's increasing subsequence problem, tiling problems,
(Aztec diamond, hexagon tiling, \dots), random particle systems (TASEP, \dots),  
\dots 
\item statistical mechanics: Ising model, XXZ spin chain, 6 vertex model,
\dots  
\item classical analysis: Riemann-Hilbert problems, orthogonal polynomials,
(modern) special function theory (Painlev\'e equations), \dots 
\item numerical analysis: QR, Toda eigenvalue algorithm, Singular value 
decomposition, \dots 
\item representation theory: representation theory of large groups
($S_\infty,\; U_\infty$ \dots), symmetric function theory, \dots 
\item algebraic geometry: Schottky problem, infinite genus Riemann surfaces,
\dots
\item transportation theory: Bus problem in Cuernavaca, Mexico, airline 
boarding, \dots \ .
\end{itemize}
The list of such intersections is long and constantly growing.  

The singular significance of KdV is just that the \tbf{first intersection}
that was observed and understood as such, was the junction of IM
with dynamical systems, and that was at the point of KdV.

How do we come to such a picture?  First we will give a precise
definition of what we mean by an integrable system.  Consider a simple
harmonic oscillator:
\begin{align}
\label{eq9}
&\dx = y\qquad, \quad  \dy = -\om^2 x\\
&\qquad x(t)|_{t=0} = x_0 \qquad, \quad y(t)|_{t=0} = y_0\nonumber
\end{align}
The solution of (9) has the following form:
\beq\label{eq10}
\begin{cases}
x(t; x_0, y_0) &= \frac{1}{\om} \;\sqrt{w^2 \, x^2_0 + y^2_0} \; \sin
\lf(wt + \sin^{-1} \lf(\frac{\om\, x_0}{\sqrt{\om^2\,x^2_0 +y^2_0}}
\rt)\rt) \\
y(t; x_0, y_0) &= \sqrt{\om^2\, x^2_0 +y^2_0}\; \cos \lf(wt
+ \sin^{-1} \lf( \frac{\om\, x_0}{\sqrt{\om^2\,x^2_0+ y^2_0}} \rt) \rt)
\end{cases}
\eeq
Note the following features of (10):
Let $\varphi: \RR^2 \to \RR_+ \times \lf(\RR/2\pi \ZZ\rt)$
$$
(\al, \beta) \longmapsto A = \frac{1}{\om} \; \sqrt{\om^2\, \al^2 +
\beta^2},\; \theta = \sin^{-1} \lf(\frac{\om\, \al}{\sqrt{
\om^2\, \al^2+ \beta^2}}\rt).
$$
Then 
$$
\varphi^{-1}: \RR_+ \times \lf(\RR/ 2\pi \ZZ\rt) \to \RR^2
$$ 
has the form 
$$
\varphi^{-1} (A, \theta) = (A \sin \theta, \; \om \, A \cos \theta)
$$
Thus (10) implies
\beq\label{eq11}
\begin{cases}
\eta (t;\eta_0) = \varphi^{-1} \lf( \varphi(\eta_0) + \vom \, t\rt)\\
\text{where } \quad\eta(t) = \lf( x(t), y(t) \rt), \eta_0 = (x_0, y_0), 
\quad \vom = (0, \om)
\end{cases}
\eeq
In other words:
\begin{subequations}\label{eq12}
\begin{align}
&\text{There exists a bijective change of variables }\quad \eta\longmapsto
\varphi(\eta) \quad \text{such that}\label{12a}\\
&\quad\qquad\eta (t, \eta_0) \;\text{ evolves according to (9)} 
\Rightarrow \label{eq12b} \\
&\hspace{1.5in} \varphi(\eta(t); \eta_0) = \varphi(\eta_0) 
+ t\;\vom\nonumber
\end{align}
i.e., in the variables $(A,\theta)= \varphi (\al, \beta)$, solutions of (9)
move linearly.
\beq\label{eq12c}
\begin{cases}
&\eta(t, \eta_0)\; \text{ is recovered from formula (11) via a map } \\
&\qquad \varphi^{-1} (A, \theta) = (A\sin \theta, \om \, A\cos \theta)\\[3mm]
& \text{ in which the behavior of $\sin \theta, \cos \theta$ is very well
understood.}\\
&\text{The same is true for  }\varphi. 
\text{ What we learn, in particular, based on this }\\
&\text{knowledge of $\varphi$ and $\varphi^{-1}$, \quad is that }\\
&\qquad \eta(t; \eta_0) \quad\text{evolves periodically in time 
with period $2\pi/\om$}
\end{cases}
\eeq
\end{subequations}
We are led to the following:

We say that a dynamical system $t\mapsto \eta(t)$ is 
\tbf{integrable} if 
\begin{subequations}\label{eq13}
\begin{align}\label{eq13a}
&\begin{cases}
&\text{There exists a bijective map } \;\varphi : \eta \mapsto \varphi(\eta)
\equiv \zeta\\
&\qquad \text{ such that $\varphi$ linearizes the system}\\
&\hspace{1in} \varphi\lf(\eta (t)\rt) = \varphi \lf( \eta (t=0)\rt)
+ \vom \, t\\
&\text{ and so }\\
&\qquad \eta \lf(t; \eta\,(t=0)\rt)= \varphi^{-1} \lf(\varphi \lf(\eta
(t=0)\rt) + \vom \, t\rt) 
\end{cases}\\[2mm]
&\hspace{2.25in}\text{AND} \nonumber\\
&\begin{cases}\label{eq13b}
&\text{The behavior of $\;\varphi,\,\varphi^{-1}\;$ are well enough 
understood  so that}\\
&\text{the behavior of $\eta\lf(t; \eta\,(t=0)\rt)$ as $t\to \infty$
is clearly revealed.}
\end{cases}
\end{align}
\end{subequations}
More generally, we say a system $\eta$ which depends on some parameters
$\eta= \eta \,(a, b, \dots)$ is \tbf{integrable} if 
\begin{subequations}\label{eq14}
\begin{align}
&\begin{cases}
&\text{There exists a bijective change of variables $\; \eta\to \zeta
=\varphi(\eta)$ such}\\
&\text{that the dependence of $\zeta$ on $a, b, \dots \ .$}\\
& \hspace{1in} \zeta \lf(a, b, \dots \rt) = \varphi \lf( \eta
\lf(a, b, \dots\rt)\rt)\\
&\text{is simple/well-understood}
\end{cases}\\
&\hspace{2.25in}\text{and}\nonumber\\
&\begin{cases}\label{14b}
&\text{The behavior of the function theory}\\
&\qquad\quad \eta \mapsto \zeta \equiv \varphi(\eta) \quad ,
\quad\zeta \longmapsto 
\eta = \varphi^{-1} (\zeta)\\
&\text{is well-enough understood so that the behavior of }\\
&\qquad\quad\eta\lf(a, b \dots\rt) = \varphi^{-1} 
\lf(\zeta \lf(a, b \dots, \rt) \rt) 
\end{cases}
\end{align}
\end{subequations}
is revealed in an explicit form as $a, b, \dots$ vary, becoming,
in particular, large or small.

Notice that in this definition of an integrable system, various sufficient
conditions for integrability such as commuting integrals, Lax-pairs, \dots,
are conspicuously absent.  A system is integrable, if you can solve it,
but subject to certain strict guidelines.  This is a return to McKean and 
Flaschka, an institutionalization, as it were, of the ``Wild West''.

According to this definition, progress in the theory of integrable systems
is made 

\centerline{\tbf{EITHER}}

by discovering how to linearize a new system
$$
\eta \to \zeta = \varphi(\eta)
$$
using  a \tbf{known} function theory $\varphi$.  For example:  Newton's 
problem of two gravitating bodies, is solved in terms of trigonometric 
functions/ellipses/parabolas---mathematical objects already well-known 
to the Greeks.  In the 19$^{th}$ century, Jacobi solved geodesic flow 
on an ellipsoid using newly 
minted hyperelliptic function theory, and so on, \dots

\centerline{\tbf{OR}}

\noi
by discovering/inventing a new function theory which linearizes the given 
problem at hand.  For example: To facilitate numerical calculations in 
spherical geometry, Napier, in the early 1700's, realized that what he needed
to do was to linearize multiplication
$$
\eta\, \tilde{\eta} \longrightarrow \varphi\lf(\eta\, \tilde{\eta}\rt)
= \varphi(\eta) + \varphi(\tilde{\eta})
$$
which introduced a new function theory --- the logarithm.
Historically, \tbf{no} integrable system has had greater impact
on mathematics and science, than multiplication!
There is a similar story for all the classical special functions, Bessel,
Airy, \dots, each of which was introduced to address a particular problem.

The following aspect of the above evolving integrability process is crucial
and gets to the heart of the Integrable Method (IM):  Once a new function
theory has been discovered and developed, it enters the \tbf{toolkit} of IM,
finding application in problems far removed from the original discovery.

Certain philosophical points are in order here.

\begin{enumerate}
\item[(i)] There is \tbf{no difference} in spirit, philosophically, between our
definition of an integrable system and what we do in ordinary life.  We try
to address problems by rephrasing them (read ``change of variables'') so we 
can recognize them as something we know.  After all, what else is a 
``precedent'' in a law case?  We introduce new words --- a new ``function 
theory'' --- to capture new developments and so extend and deepen our 
understanding.  Recall that Adam's first cognitive act in Genesis was to give
the animals names.  The only difference between this progression in
ordinary life versus mathematics, is one of degree and precision.
\item[(ii)] This definition presents ``integrability'' 
\tbf{not as a predetermined}
property of a system frozen in time.  Rather, in this view the status of a
system as  integrable depends on the technology/function theory available
\tbf{at the time}.  If an appropriate new function theory is developed, the
status of the system may change to integrable.
\end{enumerate}

How does one determine if a system is integrable and how do you integrate it?
Let me say at the outset, and categorically, that I believe there is no 
systematic answer to this question.  Showing a system is integrable is
\tbf{always} a matter of luck and intuition.

We do, however, have a \tbf{toolkit} which  one can bring to a problem at hand.

At this point in time, the toolkit contains, amongst others, the following
components:
\begin{enumerate}
\item[(a)] a broad and powerful \tbf{set of functions/transforms/constructions}
$$
\eta \to \zeta = \varphi(\eta)
$$ 
that can be used to convert a broad class of problems of interest in 
mathematics/physics, into ``known'' problems:  In the simplest cases
$\eta \to \varphi(\eta)$ linearizes the problem.
\item[(b)] \tbf{powerful techniques} to analyze $\varphi,\, \varphi^{-1}$
such that the asymptotic behavior of the original $\eta$-system can be 
inferred explicitly from the known asymptotic behavior of the 
$\zeta$-system, as relevant parameters, e.g.\ time, become large.
\item[(c)]  a \tbf{particular, versatile} class of functions, the 
Painlev\'e functions, which play the same role in modern (nonlinear) 
theoretical physics that classical special functions played in (linear)
19$^{th}$ century physics.  Painlev\'e functions form the \tbf{core of 
modern special function}, and their behavior is known with the same
precision as in the case of the classical special functions.  We note that 
the Painlev\'e equations \tbf{are themselves integrable} in the sense of
Definition~(14a). 
\item[(d)] a class of \tbf{``integrable'' stochastic models} --- random
matrix theory (RMT).  Instead of modeling a stochastic system by the 
roll of a die, say, we now have the possibility to model a whole new class
of systems by the \tbf{eigenvalues of a random matrix}.  Thus RMT plays 
the role of a \tbf{stochastic special function theory}.  RMT is ``integrable''
in the sense that key statistics such as the gap probability, or edge 
statistics, for example, are given by functions, e.g.\ Painlev\'e functions,
that describe (deterministic) integrable problems as above.  We will say
more about this later.  
\end{enumerate}

We will now show how all this works in concrete situations.  Note, however, 
by no means all known integrable systems can be solved using tools from the
IM-toolkit.  For example the beautiful system that Patrick G\'erard et al.\ 
have been investigating recently (see e.g.~\cite{GeLe}), seems to be something completely different.
We will consider various examples.  The first example is taken from 
dynamics, viz., the NLS equation.

To show that NLS is integrable, we \tbf{first} extract a particular tool from 
the toolkit --- the Riemann-Hilbert Problem (RHP):  
Let $\Sg \subset \CC$ be an oriented contour
and let $v: \Sg \to G \ell \,(n, \CC)$ be a map (the ``jump matrix'')
from $\Sg$ to the invertible $n\times n$ matrices, $v,\, v^{-1} \in L^\infty
(\Sg)$.  By convention, at a point $z\in \Sg$,  
the $(+)$ side (respectively
$(-)$ side) lies to the left (respectively right) as one traverse $\Sg$
in the direction of the orientation, as indicated in Figure 2.
Then the (normalized) RHP $(\Sg, v)$ consists in finding an $n\times n$
matrix-valued function $m=m(z)$  such that
\begin{align*}
&\bullet m(z) \quad\text{is analytic in } \CC/\Sg\\
&\bullet m_+(z) = m_-(z)\;v(z) \quad, \quad z\in \Sg\\
&\qquad\text{ where }\quad m_\pm (z) = \lim_{z' \to z_\pm} \; m(z') \\
&\bullet m(z) \to I_n \quad\text{ as }\quad z\to \infty
\end{align*}
Here ``$z'\to z_\pm$'' denotes the limit as $z' \in \CC/\Sg$ approaches 
$z\in \Sg$ from the $(\pm)$-side, respectively.  
\begin{figure}[tbp]\label{fig2}
  \centering
  \includegraphics[width=.55\linewidth]{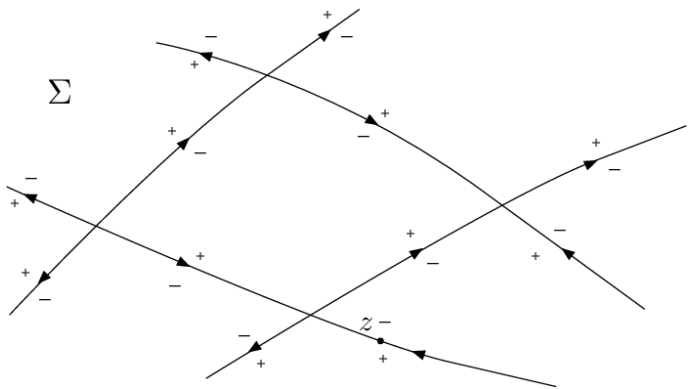}
\caption{Oriented Contour $\Sg$} 
\end{figure}
The particular contour $\Sg$ and the jump matrix $v$ 
are tailored to the problem at hand.

There are many technicalities involved here:  Does such an 
$m(z)$ exist?  In what sense do the limits $m_\pm$ exist?  And so on 
\dots \ .  Here we leave such issues aside.  RHP's play an analogous 
role in modern physics that integral representations play for 
classical special functions, such as the Airy function $Ai(z)$, Bessel
function $J_n(z)$, etc.  
For example, $Ai(z)= \frac{1}{2\pi i} \int_{\msC} \exp \lf( \frac{t^3}{3}
- z \, t\rt)dt$ for some appropriate contour $\msC \subset \CC$, which makes
it possible to analyze the behavior of $Ai(z)$ as $z\to \infty$, using the 
classical steepest descent method.

Now consider the defocusing NLS equation  on $\RR$
\beq\label{eq15}
\begin{cases}
i \, u_t + u_{xx} - 2 |u|^2\, u=0 \\
u(x,t)\Big|_{t=0} = u_0 (x) \to 0 \quad\text{as}\quad |x| \to \infty.
\end{cases}
\eeq

In 1972, Zakharov and Shabat \cite{ZaSh} showed that NLS has a Lax-pair 
formulation, as follows: Let
\begin{align*}
L(t) &= (i\, \sg)^{-1} \lf(\p_x - Q(t)\rt)\\
\intertext{where}
\sg &=  \half \begin{pmatrix}
1 & 0 \\
0 & -1
\end{pmatrix}, \qquad 
Q(t) = \begin{pmatrix}
0 & u(x,t)\\
\ovl{-u(x,t)} & 0 
\end{pmatrix}. 
\end{align*}

\bigskip
\noi
For each $t$, $L(t)$ is a self-adjoint operator acting on vector valued
function in $\lf(L^2(\RR)\rt)^2$.  Then for some explicit $B(t)$, constructed
from $u(x,t)$ and $u_x(x,t)$,
\beq\label{eq16}
u(x,t) \quad\text{solves NLS } \iff \frac{d\, L(t)}{d t} = 
\lf[B(t), L(t) \rt].
\eeq
This the \tbf{second} tool we extract from our toolkit.
So the Lax operator $L(t)$ marks a point, as it were, on our treasure map.  
How can one use $L(t)$ to solve the system?

One proceeds as follows:  This crucial step was first taken by Shabat \cite{Sha}
in the mid-1970's in the case of KdV and developed into a general scheme for
ordinary differential operators by Beals and Coifman \cite{BC} in the early 1980's.

The \tbf{map} $\boldsymbol{\varphi}$ \tbf{in 13(a)}
above for NLS is the scattering map
constructed as follows:  Suppose $u=u(x)$ is given, 
$u(x) \to 0$ sufficiently rapidly as $|x| \to \infty$.  For \tit{fixed $z\in 
\CC/\RR$}, there exists a \tit{unique} $2\times 2$ solution of the 
scattering problem
\beq\label{eq17}
L\psi = z\psi
\eeq
where
\beq\label{eq18}
m=m(x, z;\, u) \equiv \psi (x,z)\, e^{-iz,x \sg}
\eeq
is bounded on $\RR$ and 
$$
m(x,z; u) \to I \quad\text{ as } x \to -\infty.
$$
For \tbf{fixed} $\boldsymbol{x\in \RR}$, such so-called \tit{Beals-Coifman
solutions} also have the following properties:
\beq\label{eq19}
\begin{aligned}
&m(x,z; u) \;\text{ is analytic in $z$ for } \; z\in \CC/\RR\\
&\qquad \text{ and continuous in } \ovl{\CC}_+ \text{ and  in } \ovl{\CC}_-
\end{aligned}
\eeq
\beq\label{eq20}
m(x, z; u) \to I \quad\text{ as }\quad z\to \infty \quad\text{in}\quad
\CC/\RR.
\eeq
Now both $\psi_\pm (x,z; u)= \lim_{z' \to z_\pm} \;\psi(x, z;u)\;,\; z\in \RR$
clearly solve $L\,\psi_{\pm} = z\, \psi_{\pm}$ which implies that there exists
$v=v(z)=v(z;u)$ independent of $x$, such that for all $x\in \RR$
\beq\label{eq21}
\psi_+ \lf(x, z\rt)= \psi_- \lf(x,z\rt) v(z)\quad, \quad
z\in \RR
\eeq
or in terms of 
\beq\label{eq22}
m_\pm = \psi_\pm (x,z)\, e^{-ixz \sg}
\eeq
we have 
\beq\label{eq23}
m_+ (x,z)= m_- (x,z)\, v_x(z)\quad, \quad z\in \RR
\eeq
where
\beq\label{eq24}
v_x(z) = e^{ixz\, \sg} \, v(z)\, e^{-ixz\,\sg}
\eeq
Said differently, for each $x\in \RR$, $m(x,z)$
solves the normalized RHP $\lf(\Sg, v_x\rt)$ where
$\Sg= \RR$, oriented from $-\infty$ to $+\infty$, and $v$ is as above.
In this way, a RHP enters naturally into the picture introduced by the 
Lax operator $L$.

It turns out that $v$ has a special form
\beq\label{eq25}
\begin{cases}
v(z) &= \begin{pmatrix}
1-\lf|r(z)\rt|^2 & r(z)\\
 -\ovl{r(z)} &1 
\end{pmatrix}\\[4mm]
v_x(z) &= \begin{pmatrix}
1 - |r|^2 & r\, e^{ixz}\\
- \bar{r} e^{-ixz} &1
\end{pmatrix}
\end{cases}
\eeq
where $r(z)$, the \tbf{reflection coefficient}, satisfies $\|r\|_\infty <1$.
We define the map $\varphi$ for NLS as follows:
\beq\label{eq26}
u\mapsto \varphi(u) \equiv r
\eeq
Suppose $r$ is given and $x$ fixed.  To construct $\varphi^{-1}(r)$ we must
solve the RHP $(\RR, v_x)$ with $v_x$ as in (25).  If $m=m(x,z)$ is the 
solution of the RHP, then expanding at $z=\infty$, we have
$$
m(x,z) = I + \frac{\lf(m_1(x)\rt)}{z} +O \lf(\frac{1}{z^2}\rt)\quad, \qquad 
z\to \infty.
$$
A simple calculation then shows that
\beq\label{eq27}
u(x)= \varphi^{-1} (r) = -i(m_1(x))_{12}.
\eeq
Thus
$$
\varphi \leftrightarrow\; \text{ scattering map} \quad;\quad \varphi^{-1}
\leftrightarrow \text{ RHP}.
$$

Now the key fact of the matter is that 
\beq\label{eq28}
\varphi \;\text{ linearizes  NLS}.
\eeq
Indeed if $u(t) = u(x,t)$ solves NLS with $u(x,t)\bigl|_{t=0} = u_0(x)$,
then 
\begin{align}
&r(t) = \varphi (u(t)) = r(z; u_0) \, e^{-itz^2} =\varphi(u_0)(z)
\, e^{-itz^2}\label{eq29}\\
\text{or} & \qquad\qquad   \log r(t) = 
\log r (z, u_0) - i \, t\, z^2 \nonumber
\end{align}
which is linear motion!

This leads to the celebrated solution procedure
\beq\label{eq30}
u(t) = \varphi^{-1} \lf(\varphi(u_0) (\cdot) \, 
e^{-i\,t(\cdot)^2}\rt).
\eeq
Thus condition (13a) for the integrability of NLS is established.

But condition (13b) is also satisfied.  Indeed the analysis of the
scattering map $u\to r = \varphi(u)$ is classical and 
well-understood.  The inverse scattering map is \tbf{also}
well-understood because of the nonlinear steepest descent method for
RHP's introduced by Deift and Zhou in 1993 \cite{DZ1}
\footnote{This paper also contains some history of earlier approaches
to analyze the behavior of solutions of integrable systems asymptotically.}.  
This is the \tbf{third tool}
we extract from our toolkit.  One finds, for example, that as $t\to\infty$
\begin{align}
u(x,t) &= \frac{1}{\sqrt{t}} \; \al(z_0) \;\; e^{i\,x^2/4t-i\nu(z_0)\,
\log 2t}\label{eq31}\\
&\hspace{1.5in}+O\lf(\frac{\ell n\; t}{t}\rt)\nonumber\\
\intertext{where}
z_0 &= x/2t \quad,\quad \nu(z) = -\frac{1}{2\pi} \, \log \lf(1-|r(z)|^2\rt)
\label{eq32}\\
\intertext{and}
\al(z)& \text{ is an explicit function of $r$.}\nonumber
\end{align}
We see, in particular, that the long-time behavior of $u(x,t)$ is given with
the same accuracy and detail as the solution of the linear 
Schro\"odinger equation $i\,u^0_t +u^0_{xx} =0$ which can be obtained by 
applying the classical steepest descent method to the Fourier
representation of $u^0(x,t)$
$$
u^0(x,t) = \frac{1}{\sqrt{2\pi}} \int_{\RR} \widehat{u^0}(z) 
\;e^{i(xz-tz^2)}\, dz
$$
where $\widehat{u^0}$ is the Fourier transform of $u^0(x,t=0)$.  As 
$t\to\infty$, one finds
\beq\label{eq33}
u^0 (x,t) = \frac{\widehat{u^0}(z_0)}{\sqrt{2it}} \;e^{ix^2/4t} + 
o\lf(\frac{1}{\sqrt{t}}\rt).
\eeq
We see that NLS is an integrable system in the sense advertised in (13ab).
It is interesting and important to note that NLS is also integrable in the 
sense of Liouville.  In 1974, Zakharov and Manakov in a celebrated paper
\cite{ZaMa1}, not only showed that NLS has a complete set of commuting 
integrals (read ``actions''), but also computed the ``angle'' variables
canonically conjugate to these integrals, thereby displaying NLS explicitly
in so-called ``action-angle'' form.  This effectively integrates the system
by ``quadrature'' (see page~2).
The first construction of action-angle variables for an integrable PDE is 
due to Zakharov and Faddeev in their landmark paper \cite{ZaFa} on the 
Korteweg-de Vries equation in 1971.

We note that the asymptotic formula (31) (32) for NLS was first obtained by 
Zakharov and Manakov in 1976 \cite{ZaMa2} using inverse scattering techniques, 
also taken 
from the IM toolbox, but without the rigor of the nonlinear steepest descent
method.

The next example, taken from Statistical Mechanics, utilizes another
tool from the toolkit, viz.\ the theory of integrable operators, IO's.

IO's were first singled out as a distinguished class of operators by  
Sakhnovich in the 60's, 70's and the theory of such operators was then fully 
developed by Its, Izergin, Korepin and Slavnov \cite{IIKS} in the 1990's.  
Let $\Sg$ be an oriented contour in $\CC$.  We say an operator $K$ acting on 
measurable functions $h$ on $\Sg$ is \tit{integrable} if it has a kernel of
the form 
\beq\label{eq34}
K(x,y) = \frac{\sum^n_{i=1} f_i(x)\, g_i(y)}{x-y} \quad, \quad
n<\infty, \;\;x,\,y\in \Sg,
\eeq
where
\begin{align}
&f_i,\;g_i\quad \in L^\infty(\Sg),\quad\text{and }\label{3q35}\\
&\qquad Kh(x) = \int_\Sg K(x,y)\, h(y)\, dy\nonumber.
\end{align}
If $\Sg$ is a ``good'' contour (i.e.\  $\Sg$ is a 
\tbf{Carleson curve}), $K$ is bounded in $L^p (\Sg)$ for $\;1<p<\infty$.

Integral opertors have many remarkable properties.  In particular the 
integrable operators form an algebra and $(I+K)^{-1}$,
if it exists, is also integrable if $K$ is integrable.
But most remarkably, $(I+K)^{-1}$ can be computed in 
terms of a naturally associated RHP on $\Sg$.  It works like this.
If $K(x,y) = \sum^n_{i=1} f_i(x)\; g_i(y)/x-y$, then
\begin{align}
&\lf(I +K\rt)^{-1} = I +R\label{eq36}\\
\text{where }\quad & R(x,y) = \sum^n_{i=1} F_i(x)\;G_i(y)\Big/x-y
\nonumber
\end{align}
for suitable $F_i, \;G_i$.  Now assume for simplicity that $\sum^n_{i=1}
f_i(x)\:g_i(x) =0$ and let 
\beq\label{eq37}
v(z) = I -2\pi\, f(z)\, g(z)^T\quad, \quad z\in \Sg,
\eeq
where $\qquad f= \lf(f_i, \dots, f_n\rt)^T, \;\; g=\lf(g_i, \dots, 
g_n\rt)^T$

\noi
and suppose $m(z)$ solves the normalized RHP $(\Sg, v)$.
Then
\begin{align}
F(z) &=m_+(z) \, f(z)= m_-(z) \, f(z)\label{eq38}\\
\intertext{and}
G(z) &= \lf(m_+^{-1}\rt)^T\, 
g(z)=\lf(m^{-1}_-\rt)^T \, g(z)\label{eq39}
\end{align}

Here is an example how integrable operators arise.  Consider the 
$\text{spin--}\half\; XY$ model in a magnetic field with 
Hamiltonian
\beq\label{eq40}
H=-\half \sum_{\ell\, \in\, \ZZ} 
\lf( \sg^x_\ell \; \sg^x_{\ell+1} + \sg^z_\ell \rt)
\eeq
where $\sg^x_\ell,\, \sg^z_\ell$ are the standard Pauli matrices at the 
$\ell^{th}$ site of a 1-d lattice.

As shown by McCoy, Perk and Schrock \cite{McPS} in 1983, the 
auto-correlation function $X(t)$
$$
X(t) = \langle\sg^x_0(t)\, \sg^x_0\rangle_T = 
\frac{tr \lf(e^{-\beta H} 
\lf(e^{-iHt}\, \sg^x_0\, e^{i\,Ht}\rt)\sg^x_0\rt)}{tr\, e^{-\beta H}}
$$
where $\beta=\frac{1}{T}$, can be expressed as follows:
$$
X(t) =e^{-t^2/2} \,\det \lf(1-K_t\rt)
$$
Here $K_t$ is the operator on $L^2(-1,1)$ with kernel
\beq\label{eq41} 
K_t (z, z') =\varphi(z) \;\frac{\sin i t(z-z')}{\pi\,(z-z')}
\quad, \quad -1\le z, \; z'\le 1,
\eeq
and
\beq\label{eq42}
\varphi(z) = \text{tanh} \lf(\beta \;\sqrt{1-z^2}\rt) \quad, \quad
-1 < z <1.
\eeq
Observe that 
\beq\label{eq43}
K_t(z,z') = \frac{\sum^2_{i=1} f_i(z) \, g_i(z')}{z-z'}
\eeq
where
\begin{align*}
f&= \lf(f_1, f_2\rt)^T = \lf(\frac{-e^{tz} \,\varphi(z)}{2\pi \,i}\;,\;
\frac{-e^{-tz}\, \varphi(z)}{2\pi \,i}\rt)^T\\
g&=\lf(g_1, g_2\rt)^T = \lf(e^{-tz},\;-e^{tz}\rt)^T
\end{align*}
so that $K_t$ is an integrable operator.  We have
\beq\label{eq44}
v= v_t = I-2\pi i\;f\,g^T = \begin{pmatrix}
1 + \varphi(z) & -\varphi(z)\,e^{2zt}\\
\varphi(z)\,e^{-2zt} & 1-\varphi(z)
\end{pmatrix}\;,\quad z \in (-1, 1)\ .
\eeq
As
\begin{align*}
\frac{d}{dt}\;\log \det \lf( 1-K_{t} \rt) &= \frac{d}{dt} \;tr \log
(1-K_{t})  \\
&=-tr \lf( \frac{1}{1-K_t}\; \dot{K}_t \rt)
\end{align*}
we see that $\frac{d}{dt}\;\log \det (1-K_t)$, and ultimately $X(t)$, can 
be expressed via (36) (38) (39) in terms of the solution 
$m_t$ of the RHP $\lf(\sum=(-1,1), v_t\rt)$

Applying the nonlinear steepest descent method to this RHP as $t\to \infty$,
one finds (Deift-Zhou (1994) \cite{DZ2}) that
\beq\label{eq45}
X(t) =  \exp
\lf(\frac{t}{\pi}\int^1_{-1} \;\log \lf|\text{tanh}\; \beta s\rt|ds 
+ o(t)\rt)
\eeq
This shows that $H$ in (40) is integrable in the sense that key statistics
for $H$ such as the autocorrelation function $X(t)$ for the spin $\sg^x_0$
is integrable in the sense of (14ab)
$$
X(t) \overset{\varphi}{\mapsto} \;K_t \in 
\text{ integrable operators}
$$
and $\varphi^{-1}$ is computed with any desired precision using RH-steepest
descent methods to obtain (45).  Note that the appearance of the terms 
$\varphi(z)\,e^{\pm 2zt}$ in the jump matrix $v_t$ for $K_t=\varphi
(X(t))$, makes explicit the linearizing property of the map $\varphi$.

Another famous integrable operator appears in the bulk scaling limit for the 
gap probability for invariant Hermitian ensembles in random matrix 
theory.  More precisely, consider the ensemble of $N\times N$ Hermitian
matrix $\{M\}$ with invariant distribution 
$$
P_N(M)\, dM= \frac{e^{-N\, tr\, V(M)}\,dM}{\int e^{-N\,tr\,V(M)}\,dM},
$$
where $V(x) \to +\infty$ as $|x| \to \infty $ and $dM$ is Lebesgue measure
on the algebraically independent entries of $M$.
$$
\text{Set  }P_N\lf([\al,\,\beta]\rt) = \text{ gap probability } \equiv
\;\text{ Prob } \{ M \text{ has no eigenvalues in }\; [\al,\, \beta]\},
\qquad \al < \beta.
$$
We are interested in the scaling limit of $P_N\lf([\al,\, \beta]\rt)$ i.e.
$$
P(\al, \beta) = \lim_{N\to \infty} P_N \lf(\lf[
\frac{\al}{\rho_N},\;\frac{\beta}{\rho_N}\rt]\rt)
$$
for some appropriate scaling $\rho_N \sim N$.  One finds (and here RH
techniques play a key role) that
\beq\label{eq46} 
P(\al, \beta)= \det (1-K_s) \quad, \quad s= \beta-\al
\eeq
where $K_s$ has a kernel
$$
K_s(x,y) = \frac{\sin (x-y)}{\pi (x-y)} \qquad\text{acting on }\quad
L^2(0, 2s).
$$
Clearly $\quad\ds{ K_s(x,y) = \frac{e^{ix}\;e^{-iy}-e^{-ix}\;e^{iy}}{2\pi\,i
(x-y)}}$
is an integrable operator.  The asymptotics of $P(\al, \beta)$ can then 
be evaluated asymptotically with great precision as $s\to\infty$, by 
applying the nonlinear steepest descent method for RHP's to the RHP 
associated with the integrable operator $K_s$, as in the case for $K_t$ in 
(44) et seq.

Thus RMT is integrable in the sense that a key statistic, the gap 
probability in the bulk scaling limit, is an integrable system in 
the sense of (14ab):
$$
\text{Scaled gap probability }
\;P_{(\al, \beta)}\; \overset{\varphi}{\longrightarrow} \;K_s(x,y)
\in \text{ Integrable operators}
$$
\begin{align*}
&\varphi^{-1} \quad\text{ is then evaluated via the formula }\;
\det \,(1-K_s)\\
&\text{which can be controlled precisely as }\; s\to\infty.
\end{align*}
The situation is similar for many other key statistics in RMT.
It turns out that $P_{(\al, \beta)}$ solves the Painlev\'e V equation 
as a function of $s=\beta-\al$ (this is a famous, result of Jimbo, Miwa and
M\^{o}ri and Sato, 1980 \cite{JMMS}).  But the Painlev\'e V equation is a classically 
integrable Hamiltonian system which is also integrable in the sense of 
(14ab).  Indeed it is a consequence of the seminal work of the 
Japanese School of Sato et al.\ that all the Painlev\'e equations can
be solved via associated RHP's (the RHP for Painlev\'e II in particular
was also found independently by Flaschka and Newell),
and hence are integrable in the sense of (14ab) and amenable to nonlinear
steepest descent asymptotic analysis, as described, for example, in the text, 
Painlev\'e Transcendents by Fokas, Its, Kapaev and Novokshenov (2006) 
\cite{FIKN}.

There is another perspective one can take on RMT as an integrable 
system.  The above point of view is that RMT is integrable because key 
statistics are described by functions which are solutions of {classically
integrable Hamiltonian systems.  But this point of view is  
unsatisfactory in that it attributes integrability in one area
(RMT) to integrability in another (Hamiltonian systems).  Is there a
notion of integrability for stochastic systems that is intrinsic?  In 
dynamics the simplest integrable system is free motion
\beq\label{eq47}
\dot{x}=y, \quad\dot{y}=0 \qquad\Longrightarrow\qquad
x(t)= x_0 + y_0 \,t\quad, \quad  y(t) =y_0.
\eeq
Perhaps the simplest stochastic system is a collection of coins flipped 
independently.  Now, we suggest, just as an integrable Hamiltonian system 
becomes (47) in new variables, the analogous property for a stochastic 
system should be that, in the appropriate variables, it is integrable if it
just a product of independent spin flips.

Consider the scaled gap probability,
\beq\label{eq48}
P_{(\al, \beta)} = \text{Prob } \{\text{ no eigenvalues in $(\al, \beta)$}\}
=\det (1-K_s)
\eeq
But as the operator $K_s$ is trace-class and $0 \le K_s <1$, it follows that
\beq\label{eq49}
P_{\al,\beta} = \prod^\infty_{i=1} \lf(1-\la_i\rt)
\eeq
where $0\le \la_i <1$ are the eigenvalues of $K_s$.  Now imagine
we have a collection of boxes, $B_1, B_2, \dots\;$.  With each box we have
a coin: With probability $\la_i$ a ball is placed in box $B_i$, or 
equivalently, with probability $1-\la_i$ there is no ball placed in $B_i$.
The coins are independent.  Thus we see that the probability that there are
no eigenvalue in $(\al, \beta)$, is the same as the probability of no balls
being placed in all the boxes!

This is an intrinsic probabilistic view of RMT integrability.  It applies to 
many other stochastic systems.  For example,  consider 
Ulam's longest increasing subsequence problem:

{\it
Let  $\pi = \pi(1)\, \pi(2), \dots \pi(N)$ be a permutation in the 
symmetric group $S_N$.  If 
\beq\label{eq50}
i_1< i_2 < \dots < i_k \qquad\text{and}\qquad \pi(i_1) < \dots < \pi(i_k)
\eeq
we say that 
\beq\label{eq51}
\pi(i_1) \; \pi(i_2)\; \dots, \pi(i_k)
\eeq
is an increasing subsequence for $\pi$ of \tit{length $k$}.  Let 
$\ell_N(\pi)$ denote the greatest length of any increasing subsequence for 
$\pi$, e.g.\ for $N=6, \;\;\pi=315624 \in S_6$ has 
$\ell_6(\pi) =3$ and $356$, \; $254$ and $156$ are all longest 
increasing subsequences for $\pi$.  Equip $S_N$ with 
uniform measure.  Thus for $n\le N$.
\begin{align}
q_{n,N} &\equiv \text{Prob } (\ell_N \le n) 
\label{eq52}\\
& = \frac{\# \;\{\pi: \ell_N(\pi)\le n\}}{N!}
\nonumber
\end{align}
}
\begin{question}
How does $q_{n, N}$ behave as $n, \; N\to\infty$?
\end{question}
\begin{theorem}[Baik-Deift-Johansson, 1999 \cite{BDJ}]
Let $t\in \RR$ be given.  Then
\beq\label{eq53}
F(t) \equiv \lim_{N\to\infty} \;\text{Prob } \lf(\ell_N \le 
2 \,\sqrt{N}+ t\, N^{1/6}\rt)
\eeq
exists and is given by $e^{-\int^\infty_t (x-t)\, u^2(x)\,dx}$ where
$u(x)$ is the (unique) Hastings-McLeod solution of the Painlev\'e
II equation
\beq\label{eq54}
u''=2 u^3 + xu
\eeq
normalized such that
$$
u(x) \sim Ai(x) = \text{Airy function, \; as } x\to + \infty
$$
\end{theorem}
(The original proof of this Theorem used RHP/steepest descent methods.
The proof was later simplified by Borodin, Olshanski and Okounkov using the
so-called Borodin-Okounkov-Case-Geronimo formula.)

Some observations:
\begin{enumerate}
\item[(i)] As Painlev\'e II is classically integrable, we see that the 
map
$$
q_{n, N}\quad \overset{\varphi}{\longrightarrow} \quad
u^2(t) = -\frac{d^2}{dx^2} \; \log \, F(x)
$$
transforms Ulam's longest increasing subsequence problem into an integrable 
system whose behavior is known with precision.  There are many other classical
integrable systems associated with $q_{n, N}$ but that is another story
(see Baik, Deift, Suidan (2016) \cite{BDS}).
\item[(ii)] The distribution $F(t) = e^{-\int^\infty_t (x,t)\, u^2(x)\,dx}$
is the famous Tracy-Widom distribution for the largest 
eigenvalue $\la_{\max}$ of a random Hermitian matrix in the edge-scaling
limit. In other words, the length of the longest increasing subsequence
behaves like the largest eigenvalue of a random Hermitian matrix.
More broadly, what we are seeing here is an example of how RMT plays
the role of a stochastic special function theory describing a stochastic
problem from some other a priori unrelated area.  This is no different, 
in principle, from the way the trigonometric functions describe the 
behavior of the simple harmonic oscillator.  RMT is a very versatile
tool in our IM toolbox --- tiling problems, random particle systems,
random growth models, the Riemann zeta function, \dots, all the way
back to Wigner, who introduced RMT as a model for  the scattering 
resonances of neutrons off a large nucleus, are all problems whose solution 
can be expressed in terms of RMT.
\item[(iii)] $F(t)$ can also be written as
\beq\label{eq55}
F(t) = \det \lf(1-A_t\rt)
\eeq
where $A_t$ is a particular trace class integrable operator, the Airy
operator, with $0\le A_t<1$.  Thus $F(t) = \prod^\infty_{i=1} 
\lf(1-\tilde{\la}_i(t)\rt)$ where $\{\tilde{\la}_i(t)\}$
are the eigenvalues of $A_t$.
We conclude that $F(t)$, the (limiting) distribution for the length
$\ell_N$ of the longest increasing subsequence, corresponds to an integrable
system in the above intrinsic probabilistic sense.
\item[(iv)] It is of considerable interest to note that in recent work
Gavrylenko and Lisovyy (2016 \cite{GaLi})  have shown that the isomonodromic tau function
for general Fuchsian systems can be expressed, up to an explicit elementary
function, as a Fredholm determinant of the form $\det\,(1-K)$ for some suitable 
trace class operator $K$.  Expanding the determinant as a product of 
eigenvalues, we see that the general Fuchsian system, too, is integrable
in the above intrinsic stochastic sense.
\end{enumerate}

Another tool in our toolbox concerns the notion of a scattering system.  
Consider the Toda lattice in $\lf(\RR^{2n}, \;\om=\sum^n_{i=1}
dx_i \,\wedge\, dy_i\rt)$ with Hamiltonian
\beq\label{eq56}
H_T(x,y) = \half \sum^n_{i=1} y_i^2 + \sum^{n-1}_{i=1} \,e^{(x_i-x_{i+1})}
\eeq
giving rise to Hamilton's equations
\beq\label{eq57}
\dot{x} = (H_T)_y \quad, \quad \dot{y}=-(H_T)_x.
\eeq
The \tit{scattering map} for a dynamical system maps the behavior of the 
system in the distant past onto the behavior of the system in the 
distant future.  In my Phd I worked on abstract scattering
theory in Hilbert space addressing questions of asymptotic completeness
for quantum systems and classical wave systems.  When I came
to Courant I started to study the Toda system and I was amazed to learn 
that for this multi-particle system the scattering map could be 
computed explicitly.  When I expressed my astonishment to 
J\"urgen Moser, he said to me, ``But every scattering system is 
integrable!''  It took me some time to understand what he meant.  It goes
like this:

Suppose that you have a Hamiltonian system in $\lf(\RR^{2n}, \om=\sum^n_{i=1}
dx_i \wedge dy_i \rt)$ with Hamiltonian $H$, and suppose that the solution
$$
z(t) = \lf(x(t), y(t)\rt) \quad, \quad z(0)= \lf(x(0), y(0)\rt) =(x_0, y_0)
$$
of the flow generated by $H$ behaves asymptotically 
like the solutions $\hz(t)$ of free motion
with Hamiltonian 
$$
\hH (x,y) = \half \; y^2
$$
for which  
$$
\dot{x}=y, \quad\dot{y}=0\qquad\text{with}\qquad \hz(0) =
\lf(\hx_0, \hy_0\rt),
$$
yielding
$$
\hz (t) = \lf(\hx_0 +\hy_0 \,t, \;\hy_0\rt).
$$

As $z(t) \sim \hz(t)$ by assumption, we have as $t\to \infty$,
\begin{subequations}\label{eq58}
\begin{align}
x(t) &= t\py+ \px + o(1) \label{eq58a}\\
y(t)  &= \py + o(1) \label{eq58b}
\end{align}
for some $\px,\;\py$.
\end{subequations}

Write 
$$
z(t) = U_t(z(0))\quad, \quad \hz (t) = \hU_t\lf(\hz(0)\rt). 
$$
Then, provided  $o(1)=o\lf(\frac{1}{t}\rt)$ in (58b), 
\begin{align*}
W_t(z_0) &\equiv \hU_{-t}\; \circ \; U_t(z_0)\\
&= \hU_{-t} \lf(t\py + \px + o(1), \qquad \py  +o\lf(\frac{1}{t}\rt)\rt)\\
&= \lf(t\py + \px +o(1) -t \lf(\py +o\lf(\frac{1}{t}\rt)\rt), \qquad
\py +o\lf(\frac{1}{t}\rt) \rt) \\
&= \lf(\px +o(1) , \qquad \py +o(1) \rt).
\end{align*}
Thus
$$
W_\infty (z_0) = \lim_{t\to \infty} \,W_t(z_0)
$$
exists.  Now
\begin{align*}
W_t \;\circ \;U_s &= \hU_{-t}\; \circ\; U_{t+s}\\
&= \hU_s\;\circ \; W_{t+s}\ ,
\end{align*}
and letting $t\to\infty$, we obtain
\beq\label{eq59}
W_\infty \;\circ\; U_s = \hU_s \;\circ\; W_\infty
\eeq
so that $W_\infty$ is an intertwining operator between $U_s$ and 
$\hU_s$.

But clearly $W_t$ is the composition of symplectic maps, and so is
symplectic, and hence $W_\infty$  is a symplectic
map and hence $W^{-1}_{\infty}$ is symplectic.
Thus from (59) we see that
\beq\label{eq60}
U_s = W^{-1}_{\infty} \;\circ \; \hU_s \;\circ\; W_\infty
\eeq
is symplectically equivalent to free motion, and hence is integrable. 
In particular if $\{\hlm_k\}$ are the Poisson commuting integrals 
for $\hH$, then $\{\la_k = \hlm_k \;\circ\;W_\infty\}$ are the 
(Poisson commuting) integrals for $H$.

What this computation is telling us is that if  a system is scattering, 
or more generally, if the solution of one system \tit{looks}
asymptotically like some other system, then it is in fact (equivalent to)
that system.  Remember the famous story of Roy Cohn during the
McCarthy hearings, when he was trying to convince the panel that a particular
person was a Communist?  He said:  ``If it looks like a duck, walks like a
duck, and quacks like a duck, then it's a duck!''

Now direct computations, due originally to Moser, show that the 
Toda lattice is \tit{scattering} in the sense of (58(a)(b)).  And 
so what Moser was saying is that the system is \tbf{necessarily} integrable.
The Toda lattice is a rich and wonderful system and I spent much of the 
1980's analyzing the lattice and its various generalizations together
with Carlos Tomei, Luen-Chau Li and Tara Nanda.  I will say much more about
this system below.  It was a great discovery of Flaschka \cite{Fla} (and later 
independently, Manakov \cite{Man}) that the Toda system indeed had a Lax pair 
formulation (see (74) below).  

The idea of a scattering system can be applied to PDE's.  Some 15--20
years ago Xin Zhou and I \cite{DZ3} began to consider perturbations of the 
defocussing NLS equation on the line, 
\beq\label{eq61}
i\,u_t +u_{xx} -2 |u|^2\,u - \ep |u|^\ell \, u=0, \qquad \ell >2 
\eeq
with
$$
u(x,\, t=0) = u_0(x) \to 0 \quad\text{ as }\quad |x| \to \infty.
$$
In the spatially periodic case, $u(x,t) = u(x+1, \,t)$,  solutions of 
NLS (the integrable case: $\ep =0$) move linearly on a (generically infinite
dimensional) torus.  In the perturbed case $(\ep \neq 0)$, KAM methods
can be (extended and ) applied (with great technical virtuosity) to show
(here Kuksin, P\"oschel, Kappeler have played the key role) that,
as in the familiar finite dimensional case, some finite dimensional tori
persist for (61) under perturbation.  However, on the whole line with 
$u_0(x) \to 0$ as $|x| \to \infty$, the situation, as we now describe, is
very different. 

In the spirit of it ``walks like a duck'', what is the ``duck'' for 
solutions of (61)?  The ``duck'' here is a solution $\pu(x,t)$ of 
the NLS equation.
\beq\label{eq62}
\begin{aligned}
i\,\pu_t + \pu_{xx} - 2|\pu|^2 \, \pu&=0 \\
\qquad \pu(x,0)= \pu_0(x) &\to 0\qquad \text{as} \qquad|x| \to \infty.
\end{aligned}
\eeq

Recall the following calculations from classical KAM theory in $R^{2n}$, say:
Suppose that the flow with Hamiltonian $H_0$ is integrable and $H_\ep=
H_0 + \ep \hH$ is a perturbation of $H_0$.  Hamilton's equation for 
$H_\ep$ has the form
\beq\label{eq63}
z_t = J \nab H_\ep = J \nab H_0 + \ep \, J\nab \hH , \quad, \quad
z(0) =z_0
\eeq
with $J=\lf( \begin{smallmatrix} O & I_n \\ -I_n &0  \end{smallmatrix}\rt)$.  
If $\;\;J\nab H_0$
is linear in $z$, say $\;\;J\nab H_0 = Az$, then we can 
solve (63) by D'Alembert's principle to obtain
\beq\label{eq64}
z(t) = e^{At}\;z_0 + \ep \int^t_0 e^{A(t-s)}\;J\nab\hH(s) \, ds
\eeq
to which an iteration procedure can be applied.  If $J\nab H_0$ is 
\tbf{not linear}, however, \tbf{no such} D'Alembert formula exist, and 
this is the reason that the starting point for any KAM investigation
is to first write (63) in action-angle variables $z\mapsto \zeta$ for $H_0$:
Then $J \nab H_0$ is linear and (64) applies.

With this in mind, we used the linearizing map for NLS described in (26)
$$
u(x,t) = u(t) \mapsto \varphi(u(t)) =  r(t) = r(t;z)
$$
as a change of variables for the perturbed equation (61).  And although the 
map $\varphi$ no longer linearizes the equation, it does transform the 
equation into the form 
\beq\label{eq65}
\frac{\p r}{\p t} \;(t,z) = -i\,z^2\,r(t,z) + \ep \,F\lf(z,t; r(t)\rt)
\eeq
to which D'Alembert's principle can be applied
\beq\label{eq66}
\hr(t,z) = r_0(z) + \ep \int^t_0 F \lf( z,s; \hr\,
e^{-i\,s<\;\;>^2}\rt)ds
\eeq
where $r_0(z) = \varphi(z)$ and $\hr(t,z)=r(t,s)\,e^{itz^2}$.  The functional
$F$ depends on $\varphi$ and $\varphi^{-1}$, and so, in particular, 
involves the RHP $\lf(\Sg= \RR,\, v_t\rt)$.  Fortunately this RHP can 
be evaluated with sufficient accuracy  using steepest descent methods 
in order to obtain the asymptotics of $\hr(t,z)$ as $t\to \infty$, 
and hence of $u(x,t)= \varphi^{-1} \lf(\hr(t)\, e^{-it<\;\;>^2}\rt)$.

Let $U^\ep_t(u_0)$ be the solution of (61) and $U_t^{NLS} \lf(\pu_0\rt)$ 
be the solution of NLS (62) with $u_0, \pu_0$ in $H^{1,1}= \lf\{f\in L^2
(\RR): f'\in L^2(\RR), \quad x\,f \in L^2(\RR)\rt\}$, respectively.
Then the upshot of this analysis is, in particular, that 
\beq\label{eq67}
W^\pm (u_0) = \lim_{t\to \pm \infty} \quad U^{NLS}_{-t} \;\circ \;
U^\ep_t(u_0)
\eeq
exist strongly which shows that as $t\to \pm \infty$\ ,
$$
U^\ep_t (u_0) \sim U^{NLS}_t \lf(W^\pm (u_0) \rt)
$$
and much more.  In particular, there are commuting 
integrals for (61), \dots,

Three observations:
\begin{enumerate}
\item[(a)] As opposed to KAM where integrability is preserved on sets of high
measure, here integrability is preserved on open subsets of full measure.
\item[(b)] As a tool in our IM toolbox, integrability makes it possible
to analyze perturbations of integrable systems, via a D'Alembert principle.
\item[(c)] There is a Catch 22 in the whole story.  Suppose you say,
``my goal is describe the evolution of solutions of the perturbed equation
(61) as $t\to\infty$''.  To do this one must have in mind what the 
solutions should look like as $t\to \infty$: Do they look like solutions 
of NLS, or perhaps like solutions of the free Schr\"odinger equation 
$i\,u_t + u_{xx} =0$?  Now suppose you disregard any thoughts on integrability 
and utilize any method you can think of, dynamical systems ideas, etc.,
to analyze the system and you find in the end that the solution indeed behaves
like NLS.  But here's the catch; if it looks like NLS, then the wave operators
$W_\pm$ in (67) exist, and hence the system is integrable!  It looks like a 
duck, walks like a duck and quacks like a duck, and so it's a duck!  In 
other words, whatever methods you used, they would not have succeeded
unless the system was integrable in the first place!
\end{enumerate}

Finally, I would like to discuss briefly an extremely useful algebraic 
\tbf{tool} in the IM toolbox, viz., 
Darboux transforms/Backlund transformations.  These are 
explicit transforms that convert solutions of one (nonlinear)
equation into solutions of another equation, or into different solutions
of the same equation.  For example, the famous Miura transform, a
particular Darboux/Backlund transform, 
$$
v(x,t) \to u(x,t) = v_x (x,t) + v^2(x,t)
$$
converts solutions $v(x,t)$ of the modified KdV equation
$$
v_t + 6v^2 \,v_x + v_{xxx} =0
$$
into solutions of the KdV equation
$$
u_t + 6uu_x = u_{xxx} =0.
$$
Darboux transforms can be used to turn a solution of KdV without
solitons into one with solitons, etc.  
Darboux/Backlund transforms also turn certain spectral problems into other
spectral problems with (essentially) the same spectrum, for example, 
\begin{align*} 
H&=- \frac{d^2}{dx^2} + q(x) \longrightarrow \tH=-\frac{d^2}{dx^2}
+ \tq(x)\\
\text{where }&\qquad\tq = q-2\frac{d^2}{dx^2} \;\log \varphi, 
\qquad\text{and $\varphi$ is any solution of $H\varphi=0$},
\end{align*}
constructs $\tH$ with (essentially) the same spectrum as $H$.  
Thus a Darboux/Backlund transform is a basic isospectral action.
The literature on Darboux transforms is vast, and I just want to discuss
one application to PDE's which is perhaps not too well known.

Consider the Gross-Pitaevskii equation in one-dimension, 
\begin{align}
i\,u_t + \half \:u_{x_x} + V(x)\,u +|u|^2\,u=0
\label{eq68}\\
u(x,0) = u_0(x). \nonumber
\end{align}
For general $V$ this equation is very hard to analyze.
A case of particular interest is where 
\beq\label{eq69}
V(x) = q\, \de(x), \qquad q\in \RR
\quad\text{and $\de$ is the delta function}.
\eeq
For such $V$, (68) has a particular solution 
\beq\label{eq70}
u_\la (x,t) = \la\,e^{i\la^2\,t/2} \text{ sech }
\lf( \la |x| + \tanh^{-1} \lf(\frac{q}{\la} \rt) \rt)
\eeq
for any $\la> |q|$.  This solution is called the Bose-Einstein
condensate for the system.
\begin{question}
Is $u_\la$ asymptotically stable?  In particular, if 
\beq\label{eq71}
u(x,t=0)= u_\la \lf(x,t=0\rt) + \ep \, w(x)\quad,\quad\ep \quad\text{small,}
\eeq
does
$$
u(x,t)\quad \to\quad u_\la(x,t)\qquad \text{ as } \qquad t\to \infty?
$$
\end{question}

In the case where $w(x)$ is even, one easily sees that the initial value
problem (IVP) (68) with initial value given by (71) is equivalent to the 
initial boundary value problem (IBVP) 
\beq\label{eq72}
\begin{cases}
\qquad iu_t + \half \;u_{x_x} + |u|^2\, u = 0, \qquad x>0, \quad t>0\\
\qquad \qquad u(x, t=0) =\text{(71) for $\;x>0$ } \\
\text{subject to the Robin boundary condition at $x=0$}\\
\qquad \qquad u_x(0,t) + q\, u(0,t)=0. 
\end{cases}
\eeq

Now NLS on $\RR$ is integrable, but is NLS on $\{x>0\}$ 
with boundary conditions as in (72)
integrable?  Remember that the origin of the boundary condition is the physical
potential (read ``force''!) $V(x)$. 
So we are looking at a dynamical system, which is integrable on $\RR$,
interacting with a new ``force'' $V$.  It is not at all clear, a priori,
that the \tbf{combined system} is integrable in the sense of (13ab).

The stability question for $u_\la$ was first consider by Holmer and 
Zworski (2009) \cite{HoZ}, and using dynamical systems methods,  they 
showed asymptotic
stability of $u_\la$ for times of order $|q|^{-2/7}$.  But what about 
times larger than $|q|^{-2/7}$?  Following on the work of Holmer and
Zworski, Jungwoon Park and I \cite{DeP} begin in 2009 to consider this question.
Central to our approach was to try to show that the IBVP for NLS
as in (72) was integrable, and then use RH/steepest-descent methods. 
In the linear case, a standard approach is to use the method of images:
for Dirichlet and Neumann boundary conditions, one just reflects, 
$u(x) =-u(-x)$ or $u(x) =+u(-x)$ for $x< 0$, respectively.

For the Robin boundary condition in the linear case, 
the reflection is a little more complicated, but still standard.  
In this way one then gets an IVP on the 
line that can be solved by familiar methods.
In the non-linear case, similar methods work for the Dirichlet and Neumann
boundary conditions, but for the Robin boundary condition case, $q\neq 0$,
how should one reflect across $x=0$?  It turns out that there is a beautiful
method due to  Bikbaev and Tarasov where they construct a particular
Darboux transform version  $b(x)$ of the initial data 
$u(x,t=0), \; x>0$, and then define
\beq\label{eq73}
\begin{cases}
v(x) &= b(-x) \qquad\qquad x<0\\
&= u(x, t=0) \qquad x>0.
\end{cases}
\eeq
If $v(x,t)$ is the solution of (the integral equation) of NLS on $\RR$ with 
initial conditions (73), then $v(x,t)\Big|_{x>0}$ is a solution of the 
IBVP (72) for $t\ge 0$.  In other words, the Darboux transform can function 
as a tool in our toolkits to show that a system is integrable.

Applying RH/steepest descent methods to $v(x,t)$, one finds that $u_\la$ is 
asymptotically stable if $q>0$, but for $q<0$, generically, $u_\la$ is
not asymptotically stable: In particular, for times $t>>|q|^{-2}$, as
$t\to\infty$, a second ``soliton'' emerges and one has a ``two soliton''
condensate.  

We note that (72) can also be analyzed using Fokas' unified integration
method instead of the Bikbaev-Tarasov transform, as in Its-Shepelsky (2012)
\cite{ISh}.

\subsection*{Algorithms}

As discussed above, the Toda lattice is generated by the Hamiltonian
$$
H_T(x,y) = \half \,\sum^n_{i=1} \, y^2_i + \sum^{n-1}_{i=1}
\, e^{(x_i - x_{i+1})}.
$$
The key step in analyzing the Toda lattice was the discovery by Flaschka
\cite{Fla}, and later independently by Manakov \cite{Man}, that the Toda 
equations have a Lax-pair formulation
\beq\label{eq74}
\lf\{
\begin{gathered}
\frac{dx}{dt} = H_{T,y}\quad, \quad \frac{dy}{dt} =- H_{T,x}\\
\equiv\\
\frac{dM}{dt} = \lf[ M, B(M)\rt]
\end{gathered} \rt.
\eeq
where 
$$
M= \begin{pmatrix}
a_1 &b_1 & & \\
b_1 & \ddots &\ddots &0 \\
& \ddots & \ddots & \ddots\\
&0 &\ddots &&b_{n-1}\\
&&&b_{n-1} & a_n
\end{pmatrix}\quad, \qquad 
\begin{aligned}
B(M) = \begin{pmatrix}
0 & -b_1 & & \\
b_1 & 0 &-b_2 &0 \\
& b_2 & \ddots & \ddots\\
&0 &\ddots &&-b_{n-1}\\
&&&b_{n-1} &0
\end{pmatrix} = M_- - M^T_-
\end{aligned}
$$
and
\beq\label{eq75}
\begin{aligned}
a_k &= -y_k/2 \quad, \qquad\qquad 1 \le k \le n\\
b_k &= \half \;e^{\half \lf(x_k-x_{k+1}\rt)}\quad, \qquad 1\le k \le n-1.
\end{aligned}
\eeq
In particular, the eigenvalues $\{ \la_n \}$ of $M$ are constants of the 
motion for Toda, $\{\la_n(t) =\la_n, t\ge 0\}$. Direct calculation shows that
they are independent and Poisson commute, so that Toda is Liouville
integrable.  Now as $t\to \infty$, one can show, following Moser (1975), that
the off diagonal entries $b_k(t) \to 0$ as $t\to \infty$.  As noted by 
Deift, Nanda and Tomei (1983) \cite{DNT}, what this means is that Toda 
gives rise to an eigenvalue algorithm:

{\it
Let $M_0$ be given and let $M(t)$ solve the Toda equations (74) with 
$M(0) =M_0$. Then
\beq\label{eq76}
\begin{aligned}
&\bullet \qquad t\mapsto M(t) \; \text{ is  isospectral, }
\text{ spec $\lf(M(t)\rt)$} = \text{spec $(M_0)$}.\\
&\bullet \qquad M(t) \to \text{ diag } \lf(\la_1, \dots, \la_n\rt) \qquad
\text{as} \qquad t\to\infty.
\end{aligned}
\eeq
} 
Hence $\la_1, \dots, \la_n$ must be the eigenvalues of $M_0$.

Note that $H_T(M) = \half \;tr \;M^2$.

Now the default algorithm for eigenvalue computation is the $QR$ algorithm.
The algorithm without ``shifts'' works in the following way.  Let $M_0=M^T_0$,
$\det M_0 \neq 0$, be given, where $M_0$ is $n\times n$.  

Then $M_0$ has a unique QR-factorization
\beq\label{eq77}
\begin{aligned}
M_0=Q_0\, R_0 \quad, \qquad\qquad 
&\text{$Q_0$ orthog,\; $R_0$ upper triangular } \\
&\qquad \text{ with $(R_0)_{ii} >0, \quad i=1, \dots, n$}.
\end{aligned}
\eeq
Set
\begin{align*}
M_1 &\equiv R_0\, Q_0\\
&= Q^T_0\, M_0\, Q_0
\end{align*}
from which we see that
$$
\text{spec } (M_1) = \text{spec } M_0.
$$
Now $M_1$ has its own QR-factorization 
$$
M_1=Q_1\,R_1
$$
Set
\begin{align*}
M_2 &= R_1\,Q_1\\
&= Q^T_1\, M_1\, Q_1
\end{align*}
so that again $\text{spec }M_2 = \text{spec }M_1 = \text{spec }M_0$.

Continuing, we obtain a sequence of isospectral matrices
$$
\text{spec }M_k   =\text{spec } M_0 \quad, \quad k>0,
$$
and as $k\to\infty$, generically,
$$
M_k \to \text{ diag } \lf(\la_1, \dots, \la_n\rt)
$$
and again $\la_1, \dots, \la_n$ must be the eigenvalues of $M_0$.
If $M_0$ is tridiagonal, one verifies that $M_k$ is tridiagonal for all $k$.

There is the following \tit{Stroboscope Theorem} for the $QR$ algorithm
(Deift, Nanda, Tomei (1983) [DNT]), which is motivated by earlier work of 
Bill Symes [Sym]: 
\beq\label{eq78}
\text{Theorem\;($QR$: tridiagonal)}\phantom{xxxxxxxxxxxxxxxxXXXXXXXXXXXXXXXx}
\eeq

{\it Let $M_0 = M^T_0$ be tri-diagonal.  Then there exists a Hamiltonian flow
$t\mapsto M_{QR}(t)$, $\;M_{QR}(0)=M_0$ with Hamiltonian
\beq\label{eq79}
H_{QR}(M) = tr \; \lf(M\log M - M\rt)
\eeq
with the properties
\begin{enumerate}
\item[(i)] the flow is completely integrable 
\item[(ii)] (Stroboscope property) $M_{QR}(k) = M_k,\quad k\ge 0$, 
\quad where $\;M_k$ \:
are the $QR$ iterates starting at $M_0$, $\;\det \; M_0 \neq 0$
\item[(iii)] $M_{QR}(t)$ commutes with the Toda flow
\item[(iv)] $\frac{dM}{dt} = \lf[B(\log M),\; M\rt],\quad B(\log M)=
(\log M)_- - (\log M)_-^T$\;\;.
\end{enumerate}
}

More generally, for any $G:\RR \to \RR$
\begin{align*}
H_G(M) &= tr \;G(M)\\
\rightarrow \;\dM_G &=
\lf[ M, B\lf(g(M)\rt)\rt] \quad, \quad
g(M)= G'(M)
\end{align*}
generates an eigenvalue algorithm, so in a concrete sense, we can say, 
at least in the tri-diagonal case, that eigenvalue computation is an 
integrable process.

Now the Lax equation (74) for Toda clearly generates a global flow 
$t\mapsto M(t)$ for all full symmetric matrices $M_0 = M^T_0$.

\noi
Question.  (i) Is the Toda flow for general symmetric matrices $M_0$
Hamiltonian?

(ii) Is it integrable?

(iii) Does it constitute an eigenvalue algorithm i.e.\ 
spec $\lf(M(t)\rt) = \text{ spec } (M_0)$, 
$M(t) \to \text{ diagonal \;\; as \; } t\to\infty$?

(iv) Is there a stroboscope theorem for general $M_0$?

As shown in \cite{DLNT}, the answer to all these questions is in 
the affirmative.  Property (ii) is particularly novel.  The Lax-pair 
for Toda only gives
$n$ integrals, viz.\ the eigenvalues of $M(t)$, but the dimension of the
symplectic space for the full Toda is generically of 
$\text{dimension } 2\lf[\frac{n^2}{4}\rt]$, so one needs 
of order $\frac{n^2}{4} > > n$ Poisson
commuting integrals.  These are obtained in  
the following way:  consider, for example, the case $n=4$.  Then 
$\lf[\frac{n^2}{4}\rt]=4$
\begin{align*}
&\bullet \quad\det\; (M-z)=0 \qquad\text{has 4 roots}\quad \la_{01},\, 
\la_{02},\, \la_{03},\,\la_{04}\\
&\bullet\quad \det\; (M-z)_1=0 \qquad\text{has 2 roots}\quad 
\la_{11},\, \la_{12} 
\end{align*}
where $(M-z)_1$ is obtained by chopping off the $1^{st}$ row and last 
column of $M-z$
$$
\lf(
\begin{array}{llll}
x-z &x &x&\\
\hline\\[-7mm]
x &x-z &x & \vline \\
x &x &x-z &\vline  \\
x &x &x & \vline 
\end{array}
\begin{array}{l}
x\\
x\\
x\\
x-z
\end{array}
\rt)
$$
\begin{align*}
\text{Now } &\hspace{.5in} \la_{-1} + \la_{02} + \la_{03} + \la_{04} 
= \text{trace } M \phantom{xxxxxxxxxxxxxxxxxxxxxxxxxxxxxxx}\\
\text{and } &\hspace{.5in}  \la_{11} + \la_{12} = \text{``trace''  of } M_1
\end{align*}
are the co-adjoint invariants that specify the $8=10-2=2
\lf[\frac{n^2}{4}\rt]$ dimensional symplectic leaf $\mL_{c_1, c_2} =
\{ M: \,tr\, M = c_1, \;tr\, M_1 = c_2 \}$ on which the Toda flow is
generically defined.
The four independent integrals needed for integrability are then 
$\la_{01}, \la_{02}, \la_{03}, \la_{11}$. 
For general $n$, we keep chopping: $(M-z)_2$ is obtained by chopping off 
the first two rows and last two columns, etc.  The existence of these
``chopped'' integrals, and their Poisson commutativity follows from the 
invariance properties of $M$ under the actions of a tower of groups,
$G_1\subset G_2\subset \dots$\ .  This shows that group theory is also a tool
in the IM toolbox.  This is spectacularly true in the work of Borodin 
and Okshanski
on ``big'' groups like $S_\infty$  and $U_\infty$, and related matters.

Thus we conclude that eigenvalue computation in the full symmetric case is 
again an integrable process.

\bigskip
\noi
Remark.
The answer to Questions (i) \dots (iv) is again in the affirmative for 
general, not necessarily symmetric matrices $M\in M(n, \RR)$.  Here we
need $\sim \frac{n^2}{2}$ integrals \dots, but this is a whole other story
(Deift, Li, Tomei (1989) \cite{DLT}).

The question that will occupy us in the remainder of this paper is the 
following:  We have discussed two notions of integrability naturally 
associated with matrices:  Eigenvalue algorithms and random matrix theory.  
What happens if we try to combine these two notions?  In particular,
\beq\label{eq80}
\text{``What happens if we try to compute the 
eigenvalues of a random matrix?''}
\eeq

Let $\Sg_N$ denote the set of real $N\times N$ symmetric matrices.  Associated
with each algorithm $\mA$, there is, in the discrete case, such as $QR$,
a map $\varphi=\varphi_{\mA}:\Sg_N \to \Sg_N$ with the properties
\begin{itemize}
\item isospectrality: $\qquad\text{spec } \lf(\varphi_\mA (H)\rt) 
= \text{spec }(H)$
\item convergence:  \quad the iterates $X_{k+1}= \varphi_\mA(X_k), \quad
k\ge 0, \quad X_0 = H$ \;given, 

converge to a diagonal matrix $X_\infty, \quad X_k \to X_\infty
\quad \text{as} \quad k\to\infty$
\end{itemize}
and in the continuum case, such as Toda, there exists a flow 
$t\mapsto X(t) \in \Sg_N$ with the properties
\begin{align*}
&\bullet \quad\text{isospectrality } : \qquad \text{spec }
\lf(X(t)\rt) = \text{ spec } \lf(X(0)\rt)\\
&\bullet \quad \text{ convergence } : \qquad X(t) 
\text{ converges to a diag.\ matrix } \; X_\infty\; 
\text{ as }\;\; t\to \infty.
\end{align*}
In both cases, necessarily, the diagonal entries of $X_\infty$ are the
eigenvalues of $H$.

Given $\ep >0$, it follows, in the discrete case, that for some $m$ the 
off-diagonal entries of $X_m$ are $O(\ep)$ and hence the diagonal entries 
of $X_m$ give the eigenvalues of $H$ to $O(\ep)$.  The situation is 
similar for continuous flows $t\mapsto X(t)$.  Rather than running the algorithm
until all the off-diagonal entries are $O(\ep)$, it is customary to run the
algorithm with \tit{deflations} as follows:
For an $N\times N$ matrix $Y$ in block form 
$$
Y=\begin{pmatrix}
Y_{11} & Y_{12} \\
Y_{21} & Y_{22}
\end{pmatrix}
$$
with $Y_{11}$ of size $k\times k$ and $Y_{22}$ of size $(N-k) \times (N-k)$
for some $k\in \{1, 2, \dots, N-1\}$, the process of projecting
$$
Y\mapsto \text{ diag } \lf(Y_{11}, Y_{22} \rt)
$$
is called \tit{deflation}.  For a 
\beq\label{eq81}
\text{given }\;\ep >0, \text{ algorithm $\mA$, and matrix $H\in \Sg_N$}
\eeq
define the \tit{$k$-deflation} time.
\beq\label{eq82}
T^{(k)} (H) = T^{(k)}_{\ep,\,\mA}(H)\quad, \quad 1\le k \le N-1 \ ,
\eeq
to be the \tit{smallest} value of $m$ such that $X_m$, the $m^{th}$ iterate
of $\mA$ with $X_0=H$, has block form
$$
X_m = \begin{pmatrix}
X^{(k)}_{11} & X^{(k)}_{12}\\
X^{(k)}_{21} & X^{(k)}_{22}
\end{pmatrix}
$$
$X^{(k)}_{11}$ is $k\times k$, $\;X^{(k)}_{22}$ is $(N-k) \times (N-k)$ with
\beq\label{eq83}
\| X^{(k)}_{12} \| = \| X^{(k)}_{22}\| < \ep.
\eeq
The \tit{deflation time} $\;T(H)\;$ is then defined as 
\beq\label{eq84}
T(H) = T_{\ep\, \mA}(H) = \min_{1\le k \le N-1} \quad T^{(k)}_{\ep\, \mA}(H)
\eeq
If $\hk\in \{1, 2, \dots, N-1\}$ is such that
$$
T(H) = T^{(\hk)}_{\ep\,\mA} (H)
$$
it follows that the eigenvalues of $H$ are given by the eigenvalues of
the block diagonal matrix diag  $\lf(X^{(\hk)}_{11}, X^{(\hk)}_{22}\rt)$ 
to $O(\ep)$.  After, running the algorithm to time $T(H)$, the 
algorithm restarts by applying the basic algorithm (in parallel) to the
smaller matrices $X^{(\hk)}_{11}$ and $X^{(\hk)}_{22}$ until the next
deflation time, and so on.

In 2009, Deift, Menon, Pfrang [DMP] considered the deflation time
$\;T= T_{\ep\,\mA}$ for $N\times N$ matrices chosen from an ensemble 
$\msE$.
For a given algorithm $\mA$ and ensemble $\msE$ the authors computed
$T(H)$ for 5000--to 15000 samples of matrices $H$ chosen from $\msE$
and recorded the \tit{normalized deflation time}
\beq\label{eq85}
\tT(H) \equiv \frac{T(H) - <T>}{\sg}
\eeq
where $<T>$ is the sample average and $\sg^2$ is the sample variance
for $T(H)$ for the 5,000 to 15,000 above samples.  Surprisingly, the authors
found that
\beq\label{eq86}
\begin{cases}
\text{for a given $\ep$ and $N$, in a suitable scaling regime  
($\ep$ small, $N$ large),}\\
\text{the histogram of $\tT$ was \tit{universal}, } \\
\text{\tit{independent of the ensemble } $\msE$.}
\end{cases}
\eeq
In other words the fluctuations in the deflation time $T$, suitably scaled, 
were universal independent of $\msE$.

\newpage
Here are some typical results of their calculations (displayed in a form slightly different from [DMP])
\begin{figure}[htbp]
\centering
\subfigure[]{\includegraphics[width=.4\linewidth]{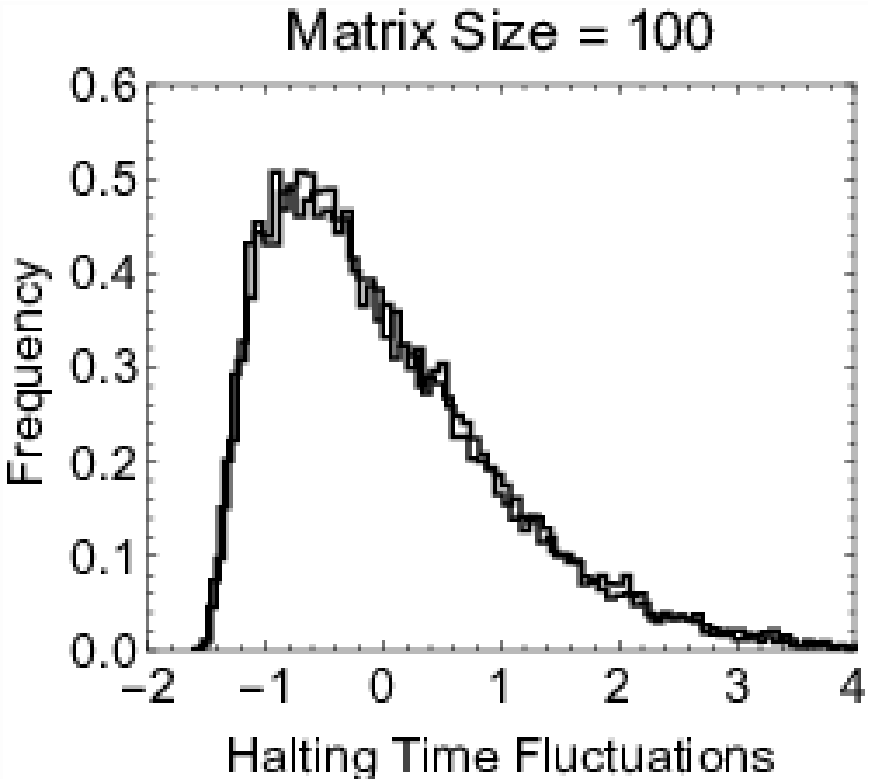}\label{f:PDM-QR}}
\hspace{.3in}\subfigure[]{\includegraphics[width=.4\linewidth]{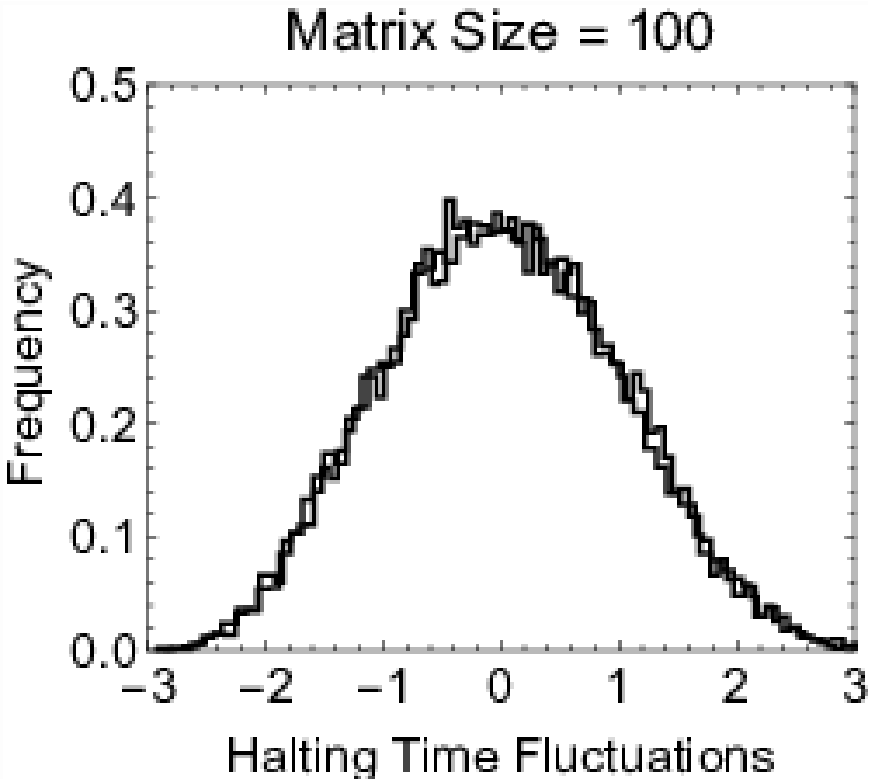}\label{f:PDM-Toda} }
\caption{\label{f:PDM} Universality for $\tilde T$ when (a) $\mathcal A$ is the QR eigenvalue algorithm and when (b) $\mathcal A$ is the Toda algorithm.  Panel (a) displays the overlay of two histograms for $\tilde T$ in the case of QR, one for each of the two ensembles $ \mathcal E = \mathrm{BE}$, consisting of iid mean-zero Bernoulli random variables and $\mathcal E = \mathrm{GOE}$, consisting of iid mean-zero normal random variables. Here $\epsilon= 10^{-10}$ and $N = 100$.  Panel (b) displays the overlay of two histograms for $\tilde T$ in the case of the Toda algorithm, and again $\mathcal E = \mathrm{BE}$ or $\mathrm{GOE}$. And here $\epsilon= 10^{-8}$ and $N = 100$.}
\end{figure}

Subsequently in 2014, Deift, Menon, Olver, Trogdon [DMOT] raised the 
question of whether the universality results of [DMP] were limited to 
eigenvalue algorithms for real symmetric matrices or whether they were 
present \tbf{more generally} in numerical computation.  And indeed, the 
authors in [DMOT], found similar universality results for a wide variety of
numerical algorithms, including
\begin{itemize}
\item other eigenvalue algorithms such as $QR$ with shifts, the Jacobi eigenvalue
algorithm, and also algorithms applied to complex Hermitian ensembles
\item conjugate gradient (CG)and GMRES (Generalized minimal residual
algorithm) algorithms to solve linear $N\times N$ systems $Hx=b$
where $b=(b_1, \dots, b_N)$ is i.i.d\ , and
\begin{align*}
H = X\, X^T, \qquad &X \text{ is $N \times m$ and random for CG}\\[-2mm]
\intertext{and}
H= I+X, \qquad &X \text{ is $N\times N$ is random for GMRES }
\end{align*}
\item an iterative algorithm to solve the Dirichlet problem
$\De u=0$ on a random star shaped region $\Om \subset \RR^2$
with random boundary data $f$ on $\p \Om$.
(Here the solution is constructed via the double layer potential method.)
\item a genetic algorithm to compute the equilibrium measure for 
orthogonal polynomials on the line
\end{itemize}
\begin{itemize}
\item a decision process investigated by Bakhtin and Correll \cite{BaCo} 
in experiments using live participants.
\end{itemize}

All of the above results were numerical/experimental.  In order to establish
universality in numerical computation as a bona fide phenomenon, and not 
just an artifact suggested, however strongly, by certain computations as 
above, it was necessary to prove universality rigorously for an algorithm
of interest.  In 2016 Deift and Trogdon \cite{DT1} considered the 1-deflation
time $T(1)$ for the Toda algorithm.  Thus one runs Toda $t\mapsto X(t)$,
$X(0)=H$, until time $t=T^{(1)}$ for which 
$$
E\lf(T^{(1)}\rt)= \sum^N_{j=2} \lf|X_{1j} \lf(T^{(1)}\rt)\rt|^2 < \ep^2 .
$$
Then $X_{11}\lf(T^{(1)}\rt)$ is an eigenvalue of $H$ to $O(\ep)$. 
As Toda is a sorting algorithm, almost surely
\beq\label{eq87}
\lf|X_{11} \lf(T^{(1)}\rt) - \la_{\max} \rt| < \ep
\eeq
where $\la_{\max}$ is the largest eigenvalue of $H$.  Thus the Toda 
algorithm with stopping time given by the 1-deflation time is an algorithm
to compute the largest eigenvalue of a real symmetric (or Hermitian) matrix.

Here is the result in \cite{DT1} for $\beta=1$ (real symmetric case) 
and $\beta=2$ (Hermitian case).  Order the eigenvalues of a real symmetric 
or Hermitian matrix by $\la_1 \le \la_2 \le \dots, \la_N$.  Then 
\beq\label{eq88}
F^{\text{gap}}_\beta(t) \equiv \lim_{N\to\infty} \text{ Prob } 
\lf( \frac{1}{C_V \, 2^{-2/3}\, N^{2/3}\lf(\la_N-\la_{N-1}\rt)}
\le t \rt) \quad, \quad t\ge 0
\eeq
exists and is universal for a wide range of invariant and Wigner ensembles.
$F^{\text{gap}}_{\beta}(t)$ is clearly the distribution function of the 
inverse of the top gap $\la_N -\la_{N-1}$ in the eigenvalues.  Here
$C_V$ is an ensemble dependent constant.

\begin{theorem}[Universality for $T^{(1)}$]
Let $\sg>0$ be fixed and let $(\ep, N)$ be in the scaling region 
$$
\mL \equiv \frac{\log \ep^{-1}}{\log N} \ge \frac{5}{3} + \frac{\sg}{2}
$$
Then if $H$ is distributed according to any real $(\beta=1)$ or 
complex $(\beta=2)$ invariant or Wigner ensemble, we have
$$
\lim_{N\to \infty} \text{ Prob } \lf( \frac{T^{(1)}}{C^{2/3}_V\;2^{-2/3}\;
N^{2/3} \lf(\log \,\ep^{-1} - 2/3 \,\log N\rt)} \le t\rt)
= F^{\text{gap}}_{\beta} (t).
$$
Thus $T^{(1)}$ behaves statistically like the inverse gap $\lf(\la_N-\la_{N-1}
\rt)^{-1}$ of a random matrix.  
\end{theorem}
Now for $\lf(\ep,\, N\rt)$ in the scaling 
region, $\;N^{2/3} \lf(\log \ep^{-1}- \frac{2}{3}\, \log N\rt)=N^{2/3}\, \log
N \lf(\al-2/3\rt)$, 
and it follows that $\text{Exp } \lf(T^{(1)}\rt) \sim N^{2/3} \, \log N$.
This is the first such precise estimate for the stopping time for an 
eigenvalue algorithm:  Mostly estimates are in the form of upper bounds,
which are often too big because the bounds must take worst case scenarios
into account.

\begin{notes}
\begin{itemize}
\item The proof of this theorem uses the most recent results on 
the eigenvalues and eigenvalues of invariant and Wigner ensembles by 
(Yau, Erd\"{o}s, Schlein, Bourgade \dots, and others (see e.g.~\cite{EY}).  
\item Similar universality results have now been proved 
(Deift and Trogdon (2017) \cite{DT2}) for 
QR acting on positive definite matrices, the power method and the inverse
power method.
\item The theorem is relevant in that the theorem describes what is 
happening for ``real life'' values of $\ep$ and $N$.  For example,
for $\ep = 10^{-16}$ and $N\le 10^9$, we have 
$\frac{\log \ep^{-1}}{\log N} \ge  \frac{16}{9} > \frac{5}{3}$.
\item Once again RMT provides a stochastic function theory to describe an
integrable stochastic process, viz., 1-deflation.  But the reverse is also
true.  Numerical algorithms with random data, raise new problems and challenges
within RMT!
\end{itemize} 
\end{notes}
\newpage

\end{document}